\documentclass[prb,aps,twocolumn,superscriptaddress,showpacs,amsmath,amssymb,floatfix]{revtex4}

\usepackage{graphicx}

\newcommand{\be}{\begin{equation}}
\newcommand{\ee}{\end{equation}}

\newcommand{\eq}[1]{(\ref{#1})}

\begin{document}
\title{Absence of Long-Range Coherence in the Parametric Emission 
from Photonic Wires}
\author{M. Wouters}
\affiliation{BEC-CNR-INFM and Dipartimento di Fisica, Universit\`a di Trento, 
I-38050 Povo, Italy}
\affiliation{TFVS, Universiteit Antwerpen, Universiteitsplein 1,
2610 Antwerpen, Belgium}
\author{I. Carusotto}
\affiliation{BEC-CNR-INFM and Dipartimento di Fisica, Universit\`a di Trento, 
I-38050 Povo, Italy}
\begin{abstract}
We analytically investigate the spatial coherence properties
of the signal emission from one-dimensional optical parametric oscillators.
Because of the reduced dimensionality, quantum fluctuations are
able to destroy the long-range phase coherence even far above
threshold.
The spatial decay of coherence is exponential and, for realistic parameters of 
semiconductor photonic wires in the strong exciton-photon coupling regime, it is
predicted to occur on an experimentally accessible length scale.
\end{abstract}


\maketitle

\section{Introduction}

A central concept in the theory of nonlinear dynamical systems is 
the so-called pattern formation phenomenon, where a spatially ordered 
structure appears in an otherwise homogeneous system when this is driven 
far from thermodynamical equilibrium~\cite{cross}.
Very important examples of pattern formation are found in the transverse dynamics of 
nonlinear optical systems~\cite{lugiato-rev,staliunas-book}, e.g. lasers, bistable devices, as well as planar  
optical parametric oscillators (OPOs).

In OPOs, the pattern formation phenomenon manifests itself as the appearance of two additional coherent beams,
called the signal and the idler, originating from nonlinear conversion of the pump beam
 when its intensity is brought beyond a threshold value~\cite{walls}.
An interesting point of view on OPO operation is provided by the 
analogy with the Bose-Einstein condensation phenomenon of equilibrium statistical 
mechanics~\cite{book}: as it happens for the matter Bose field in the 
condensate mode when the temperature is lowered below the 
condensation temperature, the population of the signal/idler 
modes becomes macroscopic above the threshold and the field becomes coherent.
Although many aspects of the OPO operation are successfully interpreted in terms
of concepts of equilibrium statistical mechanics~\cite{gatti,iac-prb},
care has to be paid as a non-equilibrium steady state is fundamentally different 
from a thermodynamical equilibrium state: it is in fact determined by 
a dynamical equilibrium between external driving and dissipation and does not follow 
a simple thermal Boltzmann law~\cite{noneqstatmech,domb}.

At equilibrium, it is well known that
long-wavelength fluctuations of both thermal and quantum origin can
destroy long-range phase coherence in reduced dimensionality Bose systems,
thus replacing true Bose-Einstein condensates with {\sl quasi}-condensates~\cite{book,popov,castin}.
First studies for the non-equilibrium case have been recently published in an optical
context~\cite{iac-prb,zambrini,staliunas} pointing out several analogies with 
the equilibrium case, but the peculiarities of the non-equilibrium case have 
not been fully appreciated yet nor experimentally investigated.

The aim of the present paper is to obtain analytical predictions
  for the long-range coherence properties of the parametric emission
  from one-dimensional photonic wires. 
Although the theory that we develop for reduced-dimensionality effects
  in non-equilibrium systems is a general one and can be extended to a
  large class of physical systems showing non-equilibrium phase
  transitions and pattern formation, our attention will be focussed on a
  specific example of solid-state system in which the effects we
  predict are experimentally accessible with the current
  semiconductor technology.

\section{The physical system \label {sec:phys}}

We consider planar semiconductor microcavities in the strong-coupling
regime~\cite{review1,review2}, a system in which OPO operation
has been recently observed~\cite{stevenson,houdre}. 
The extremely large value of the excitonic nonlinearity is responsible
for the ultra-low OPO threshold of these systems, which suggests an
enhanced role of quantum fluctuations. 
By nanostructuring planar cavities, reduced dimensionality {\em
  photonic wires} can be obtained without 
spoiling the strong-coupling condition, nor the possibility of OPO
operation~\cite{dasbach}. 
All these facts make these systems an ideal candidate for the study of
quantum effects on pattern formation in non-equilibrium systems, in
particular low-dimensional ones. 

\begin{figure}[htbp]
\begin{center}
\includegraphics[width=\columnwidth,angle=0,clip]{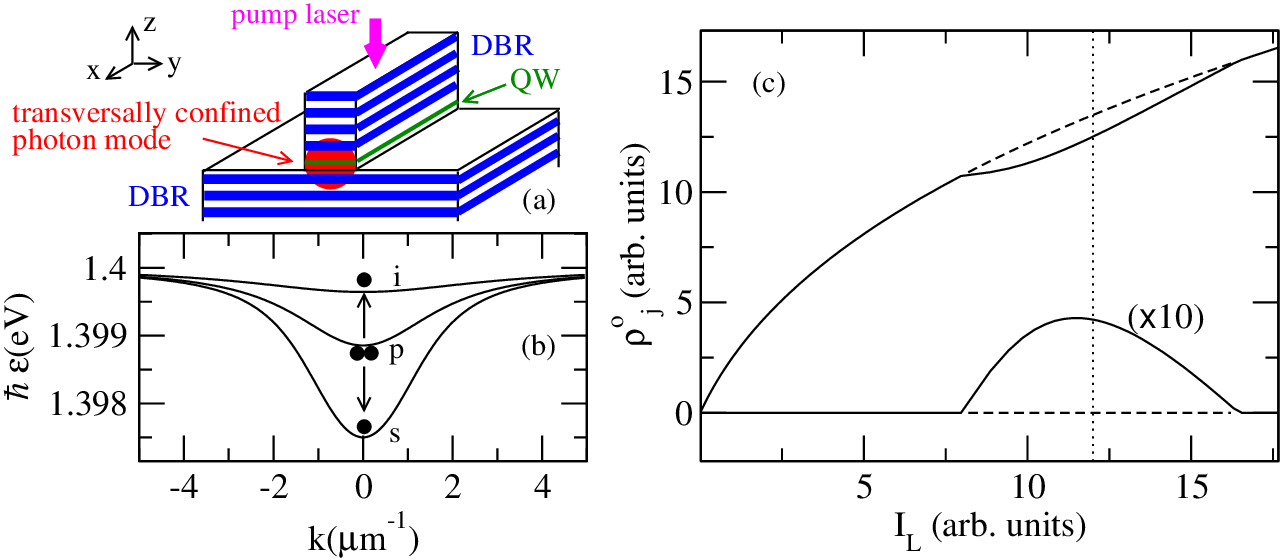}
\caption{
(a) Sketch of a semiconductor photonic wire with an embedded quantum well (QW). 
Confinement in the $z,y$ directions is provided by respectively a pair of Distributed Bragg 
Reflectors (DBR) and a lateral etching of the cavity.
(b) Linear regime  dispersions of signal, pump and idler subbands and sketch of the 
parametric process under examination.
(c) Exciton density in the pump and signal modes as a function of the pump laser intensity
for a given $\hbar \omega_p=1.398725$ eV. Full (dashed) lines represent 
linearly stable (unstable) solutions.
Cavity parameters inspired by Ref.\onlinecite{dasbach}:
$\hbar \omega^o_C=\hbar \omega^o_X=1.4$eV, $k_z=20 \mu \rm m^{-1}$, $\hbar \Omega_R=2.5$ meV, 
$L_y=4\mu$m, $\hbar\,g=5\times 10^{-6} {\rm eV} \mu \rm m^{2}$ and 
$\hbar \gamma_{s,p,i}=0.15$ meV.
}
\label{fig:MF}
\end{center}
\end{figure}  
A schematic plot of a photonic wire is given in
Fig.\ref{fig:MF}a: light is confined in the growth $z$ direction and
in the transverse $y$ direction by respectively the DBR mirrors and
the lateral etching of the structure.
On the other hand, it is free to propagate along the longitudinal
$x$ direction.
The photonic states are then classified~\cite{dasbach} by the
longitudinal wave 
vector $k$, the subband index $j$ and the polarization index
$\sigma=\{\parallel,\perp\}$, and can be used as a basis over which to expand
the (vectorial) electric field operator:
\begin{equation}
{\hat {\bf
  E}}(x,y,z)=\sum_{j\sigma}\,\int\!\frac{dk}{2\pi}\,{\hat a}_{kj\sigma}\,{\bf  
  E}_{kj\sigma}(y,z)\,e^{ikx}.
\label{electric}
\end{equation}
The polarization and spatial profile of the mode ${\bf
  E}_{kj\sigma}(y,z)$ depends on the details of the specific 
structure and has to be computed by solving Maxwell's equations for
  the specific geometry under consideration. Generally, it will show a
  non-trivial spatial and polarization structure~\cite{Panzarini}.
As a function of the longitudinal wavevector $k$ the dispersion of the
$j\sigma$ subband is approximately given~\cite{dasbach} by
\begin{equation}
\omega_{C,j\sigma}(k)=\omega_{C,j\sigma}^0\,
\sqrt{1+(k^2+k_{y,j\sigma}^2)/k_{z,j\sigma}^2},
\label{phot_disp}
\end{equation}
where $k_{y,j\sigma}$ and $k_{z,j\sigma}$ are the quantized photon
wavevectors in respectively the $y$ and $z$ directions. 

The strong dipole coupling between cavity photons and the quantum well
exciton results in a Rabi frequency $\Omega_R$ larger than the damping
rates so that the eigenmodes of the system at linear regime are
polaritons, i.e. coherent superpositions of cavity-photonic and
excitonic modes.
Compared to the photonic one \eq{phot_disp}, the dispersion of the
exciton is negligible, so that the excitonic response can be safely
considered as being local.
As a consequence, each photonic subband only mixes with the excitonic
state of wavefunction ${\bf E}_{kj\sigma}(y,z=z_{QW})$, $z_{QW}$ being
the position of the quantum well 
\footnote{This is exact as long as the confinement in
the $z$ direction is stronger than the one in the $y$ direction and
the leakage of the photonic wavefunction in the air surrounding the
structure can be neglected. 
For typical confinement sizes $L_y$ of the order of a few microns,
these approximation are very accurate~\cite{Panzarini}.}.

An example of the polaritonic dispersion $\varepsilon_j(k)$ of the
three lowest $s,p,i$ subbands is plotted in Fig.\ref{fig:MF}b
for realistic experimental parameters.
For each subband, only a single polarization state is considered here:
as experimentally shown in Ref.~\onlinecite{dasbach}, the splitting
between the $\sigma=\parallel,\perp$ polarization states in photonic
wires is in fact large enough for the parametric process to select a
single polarization state, while the other one can be safely
neglected.

Polaritons are injected into the cavity by means of a coherent and
monochromatic pump laser incident on the cavity at a frequency
$\omega_L$ close to resonance with the central $p$ subband.
In order to exclusively inject $k=0$ polaritons into the $p$ subband,
the laser field amplitude has to be constant along the $x$ direction
and to have a negligible overlap in the $y$ direction with all
subbands but for the $p$ one. 
A possible choice to fulfill both conditions, is to use a wide
laser beam incident on the sample along the $yz$ plane at an angle
chosen in such a way to match the quantized $k_y$ wavevector of the
$p$ subband.

Because of the nonlinearity due to exciton-exciton interactions within
the quantum well, two {\em pump} polaritons in the $p$ subband  
are then parametrically converted into one {\em signal} polariton in
the $s$ subband and one {\em idler} polariton in the $i$ subband.

\section{The theoretical model}
As discussed in detail in Ref.\onlinecite{iac-prb}, a simple, yet quantitative
description of the dynamics of the polariton quantum field in a microcavity 
is based on a Wigner representation of the quantum fields in terms of 
classical, yet {\em stochastic} {\bf C}-number polariton fields.
As the parametric dynamics is concentrated in the three $s,p,i$
subbands, it is useful to project the polariton field onto these
modes, so to restrict our analysis to three one-dimensional
  polaritonic fields $\phi_{s,p,i}(x)$.  
Within the Wigner framework
\footnote{The Wigner approach gives accurate results provided the
  weak-interaction condition $\gamma_j \gg
  g\,k_{\rm UV}/L_y$ is satisfied, $k_{\rm UV}$ being the UV cut-off of
  the theory, to be taken larger than any characteristic wavevector in
  the problem.}, 
the time-evolution of the fields
$\phi_{s,p,i}(x)$ is described by stochastic differential equations,
  whose form in Fourier space is: 
\begin{multline}
  i\,d \phi_j(k)= \Big\{
  \big[\varepsilon_j(k)-\frac{i}{2}\gamma_j(k)\big]
\phi_j(k)
 + \\ +2\pi\,\delta_{jp}\, \delta(k)\, F_{L} \, e^{-i\omega_L t}+\\
+\sum_{j'}\!\! \int\! \frac{dk'}{2\pi} \,\mathcal E_{jj'}(k,k')
 \phi_{j'}(k')\Big\} dt + \\
 + \sqrt{\frac{\gamma_j(k)}{4}} dW_j(k,t).
\label{mot_eq}
\end{multline}
As discussed in the previous section, the incident laser coherently
drives the pump subband $p$ with an amplitude $F_L$.
The damping rates $\gamma_j(k)$ accounts for radiative and non-radiative losses
of both the excitonic and photonic components of the polaritons.
In the region of interest
\footnote{
The $k$-dependence of the final density of states and of the tunneling
matrix element through the DBR cavity mirrors can be safely
neglected as long as the characteristic wavevectors $k_{y,j}$ and $1/\ell_c$
of the in-cavity polaritons are well inside the DBR reflection
window.}, they have a weak dependence on $k$.
In agreement with the fluctuation-dissipation
theorem~\cite{QuantumNoise}, damping is intimately connected  
to quantum fluctuations. These show up in the Wigner formalism 
as Gaussian, complex-valued, white noise terms $dW_j(k,t)$ satisfying
\begin{eqnarray}
\overline{dW_j(k,t)\,dW_{j'}(k',t)}&=&0 \\
\overline{dW_j(k,t)\,dW^*_{j'}(k',t)}&=&4\pi \,dt\,\delta(k-k')\,\delta_{jj'}.
\label{noise-terms}
\end{eqnarray}
Expectation values of symmetrically-ordered observables are then
obtained as averages of the corresponding ${\mathbf C}$-number
quantity over the noise $dW_j(x,t)$.

The nonlinear coupling responsible for the parametric process comes
from the collisional exciton-exciton interactions in the quantum
well. All other nonlinear processes are in fact much weaker and do not
contribute in a significant way to the dynamics of the system. 
At the simplest level, exciton-exciton collisional interactions can be
modelled by a polarization-independent, repulsive contact potential
with strength $g>0$~\cite{review1,review2}. 

Under the hypothesis that the polaritonic dynamics is restricted to the $s,p,i$
subbands of the lower polariton, the mean-field polariton-polariton
interaction energy can be written as 
\begin{multline}
\mathcal E_{jj'}(k,k')= 
\sum_{r,s} \!\int\! \!\frac{dp}{2\pi}\, 
g_{1D}^{jrj's} \phi_{r}^{*}(p-k) \phi_{s}(p-k'),
\label{eq:Emf}
\end{multline}
where the overlap integrals of the excitonic component of the transverse wave 
functions of the different subbands are defined as:
\begin{equation}
g^{jrj's}_{1D}=
g \,\int dy\,\phi_{X\perp}^{j*}(y)\phi_{X\perp}^{r*}(y)\phi_{X\perp}^{j'}(y)\phi_{X\perp}^{s}(y).
\label{couplingconst}
\end{equation}
The transverse excitonic wavefunctions $\phi_{X\perp}^j(y)$ are
proportional to ${\bf E}_{kj\sigma}(y,z=z_{QW})$ and normalized to the
excitonic Hopfield weight of the subband: 
\begin{equation}
\int\!dy\,|\phi^j_{X\perp}(y)|^2=|U_X^j|^2.
\end{equation}
For the sake of notational simplicity, we have not explicitely indicated
the $k$-dependence of the transverse wavefunction
$\phi_{X\perp}^j(y)$: given the small range of wavevectors $k$
involved in the parametric emission (of the order of the inverse
coherence length $1/\ell_c$), it can be safely neglected in the
following. 

The generalization of (\ref{couplingconst}) to the case of spin-dependent
exciton-exciton interactions is straightforward: in this case,
the scalar coupling constant $g$ has to the replaced by a four-indices
tensor~\cite{spinor_inter,spinor_polar} which is to be contracted with the
four (vectorial) excitonic wavefunctions $\phi_{X\perp}^{j}(y)$.

\section{Mean-field}

The {\em mean-field} solution of the motion equations (\ref{mot_eq})
once the noise terms are neglected has the simple plane wave
structure: 
\begin{equation}
\phi^o_j(x)= \sqrt{\rho^o_j} \,e^{i(k_j x- \omega_j t+\theta_j)}.
\label{pu-si}
\end{equation}
At low pump intensities, a stable solution with a finite density in the
pump mode only (i.e. $\rho_s^o=\rho_i^o=0$) exists.
The wavevector $k_p=0$, as well as the 
frequency $\omega_p=\omega_L$ are fixed by the external laser; in
particular, we have chosen $\omega_L$ to be just below the bottom of
the $p$ subband so as to avoid complicated pump bistability
effects~\cite{iac-superfl,whittaker,threshold}.
For increasing pump powers,
this solution eventually becomes dynamically unstable above some threshold, and
is replaced by a new solution with finite signal and idler intensities $\rho_{s,i}^o$.
The pump frequency and wavevector remain the same $\omega_p=\omega_L$
and $k_p=0$, while many values are possible for the signal/idler
frequencies $\omega_{s,i}$ and wavevectors $k_{s,i}$, with the only
constraint that $\omega_i=2\omega_L-\omega_s$ and $k_i=-k_s$ and that
the solution is dynamically stable.
This latter condition defines a band~\cite{cross} of stable $k_s$ values
and for each of them, the densities $\rho^o_{s,p,i}$ and the
frequencies $\omega_{s,i}$ can be simply obtained by substituting
the ansatz (\ref{pu-si}) into the deterministic part of the 
motion equation (\ref{mot_eq}).
While the phase $\theta_p$ of the pump field is fixed by the one of
the incident laser, the signal and idler ones $\theta_{s,i}$ remain
free, with only a constraint on their sum $\theta_s+\theta_i$.
The phase rotation $\theta_{s,i}\longrightarrow \theta_{s,i}\pm
\Delta\theta$ is in fact a symmetry of the problem, a $U(1)$ symmetry
which is spontaneously broken~\cite{goldstone}  above the parametric
threshold by the solution \eq{pu-si}.

The results for the signal, pump, and idler densities $\rho^o_{s,p,i}$ 
are plotted in Fig.\ref{fig:MF}c as a function of the pump laser
intensity $I_L=|F_L|^2$ for the specific case of 
a signal wavevector $k_s=0$, which is the first to become unstable
for the chosen parameters. The results would however be
qualitatively similar if a finite $k_s\neq 0$ was taken.
The signal/idler densities $\rho_{s,i}^o$ continuously
grow from zero starting from the lower threshold, get to their maximum value,
and then go back to zero at the upper threshold,
where we are back to a pump-only mean-field
solution $\rho_s^o=\rho_i^o=0$~\cite{whittaker,threshold}.

\section{Bogoliubov theory}

The simplest way to include the effect of noise terms is to linearize the
Wigner equation (\ref{mot_eq})  around the
mean-field solution (\ref{pu-si}) and treat the noise as a perturbation.
This is best done using the same ansatz:
\be
\phi_j(x,t)=\sqrt{\rho_j^o+\delta \rho_j(x,t)}\,e^{i(k_j x-\omega_j
  t+\theta_j+\delta\theta_j(x,t))}.
\label{psi-dp}
\ee
as in the generalized density-phase Bogoliubov theory
developed for the study of quasi-condensates at
equilibrium~\cite{popov,castin}.
This approach does not require the existence of
long-distance coherence, but only that density fluctuations are small
$|\delta\rho_j|\ll\rho_j^o$, and the signal and idler phases are well
defined and locally locked by the parametric process $|2\delta
\theta_p-\delta \theta_s-\delta \theta_i|\ll 1$.

Simple Ito manipulations of \eq{mot_eq}, lead to the linearized motion equation for the   
fluctuations in frequency space:
\be
\omega\, \mathcal U(k,\omega) = \mathcal L(k) \mathcal U(k,\omega) 
+ \sqrt{\Gamma(k)/4}\hspace{0.1cm} \mathcal W(k,\omega).
\label{eqlin} 
\ee
The fluctuation 6-vector is defined as 
$\mathcal U(k,\omega)=[\mathbf u^{(+)}(k,\omega), \mathbf u^{(-)}(k,\omega)]^T$,with the 3-vector:
\begin{equation}
\mathbf u^{(\pm)}_m(k,\omega) = \frac{\delta \rho_m(k,\omega)}{2 \sqrt{\rho^o_m}}
\pm i \sqrt{\rho_m^o} \,\delta \theta_m(k,\omega).
\label{w-rho-phi}
\end{equation}
The index $m$ takes here the values $1,2,3$ corresponding respectively to the subbands $s,p,i$.
The complex-valued noise 6-vector is  analogously  defined as
$\mathcal W(k,\omega)=[\mathbf w(k,\omega),-\mathbf w^*(-k,-\omega)]^T$, and is such that
$\langle \mathbf w^*_m(k,\omega)\mathbf w_n(k',\omega')\rangle=
8\pi^2\delta_{mn}\,\delta(k-k')\,\delta(\omega-\omega')$ 
and $\langle \mathbf w_m(k,\omega) \mathbf w_n(k',\omega')\rangle=0$.
The diagonal $6\times6$ damping matrix $\Gamma(k)$ has 
$\Gamma_{ll}(k)=\gamma_l(k_l+k)$ for $1 \le l \le 3$ and
$\Gamma_{ll}(k)=\gamma_{l-3}(k_{l-3}-k)$ for $4\le l \le 6$, while all other elements vanish.
The $6\times 6$ matrix $\mathcal L(k)$ has the typical Bogoliubov structure
\be
\mathcal{L}(k) =\left( 
\begin{array}{cc}
M( k) & Q \\ 
-Q^{\ast } & -M^*(-k) 
\end{array}%
\right),
\label{structL}
\ee
in terms of the $3\times 3$ matrices $M(k)$ and $Q$ whose elements (for $1 \leq m,n \leq 3$) are given by 
\begin{multline}
M_{mn}(k)=[ \varepsilon_m (k_m+k)-\omega_m-i\gamma_m(k_m+k)/2]\,\delta_{m,n}\\ 
+2 \sum_{rt}\, g_{1D}^{mnrt}\,\delta_{m,n+r-t}\, e^{i(\theta_r-\theta_t-\theta_m+\theta_n)}
\sqrt{\rho^o_{r} \rho^{o}_{t}} \\
Q_{mn}= \sum_{rt}\,{g}_{1D}^{mnrt}\,\delta_{m+n,r+t}\, 
e^{i(\theta_r+\theta_t-\theta_m-\theta_n)}
\sqrt{\rho^o_r \rho^o_t}\,.\notag
\end{multline}

\begin{figure}[htbp]
\begin{center}
\includegraphics[width=\columnwidth,angle=0,clip]{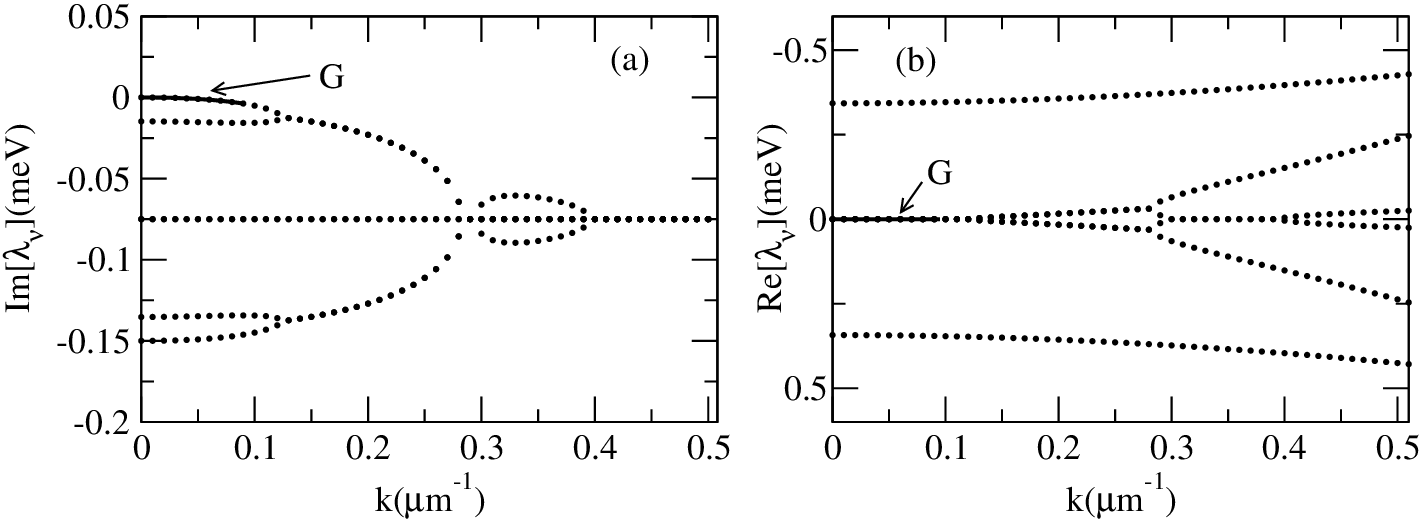}
\caption{
Imaginary (a) and real (b) parts of the
Bogoliubov excitation spectrum around the mean field steady-state. 
Solid line indicates the Goldstone branch.
Pump-intensity as at the dotted line in
Fig.\ref{fig:MF}b. Same cavity parameters as in Fig.\ref{fig:MF}.
}
\label{fig:Bogol}
\end{center}
\end{figure} 
The eigenvalues of the $\mathcal{L}(k)$ matrix give the excitation spectrum $\lambda_{\nu}(k)$ 
of our system around the mean-field steady-state solution. 
An example of such a spectrum is plotted in Fig.\ref{fig:Bogol}.
Dynamical stability of the mean-field solution requires that for all
branches $\textrm{Im}[\lambda_{\nu}(k)]<0$ for all values of $k$.
As the mean-field solution \eq{pu-si} spontaneously breaks the $U(1)$
signal/idler phase symmetry $\theta_{s,i}\rightarrow \theta_{s,i}\pm\Delta\theta$
of the motion equation \eq{mot_eq}, a Goldstone branch is necessarily present. 
For clarity, it will be denoted in the following by the index $\nu=G$.
Its main property is that its dispersion  $\lambda_G(k)$ at $k=0$ is exactly $\lambda_G(k=0)=0$ \cite{cross}.
The behavior of this eigenvalue $\lambda_G(k)$ in the neighborhood
of $k=0$ is crucial for the long-distance properties of our system.
Dynamical stability and analyticity~\cite{cross,Kato} arguments guarantee that
for $k\rightarrow 0$ one can expand
\begin{equation}
\text{Im}[\lambda_G(k)]\simeq -k^2/2a,
\label{def_a}
\end{equation}
with a strictly positive coefficient $a>0$:
differently from zero-sound in equilibrium Bose gases, the Goldstone mode 
is here a diffusive, non-propagating mode~\cite{goldstone,szymanska}.

Note that this result, being based on symmetry arguments, does
  not depend on the restriction of our model to a single polarization
  state per subband.
Explicite inclusion of all polarization states would simply increase
  the dimension of ${\mathcal L}$ to $12\times 12$.
More Bogoliubov branches would then appear in fig.\ref{fig:Bogol} but no
  significant changes would affect the nature nor the dispersion
  relation of the Goldstone mode, which (as we shall see in the next 
  section) is the only responsible for the long-distance coherence
  properties of the emission.

\section{First order correlation function}
The first-order spatial coherence of the parametric emission is
determined by the spatial coherence of the in-cavity
polaritons~\cite{iac-prb,baas} (added ref. to baas).  
In the Wigner representation, this can be written for $x_1\neq x_2$ as:
\be
g^{(1)}_s(x_1-x_2)=\big\langle \phi^*_s(x_1,t) \phi_s(x_2,t) \big\rangle_W.
\ee
Neglecting in \eq{psi-dp} the density fluctuations $\delta\rho_j$ which do not  
contribute to long-distance properties, and expanding higher order
correlations by means of the Wick theorem for Gaussian stochastic
variables~\cite{QuantumNoise}, one obtains:
\begin{equation}
g^{(1)}_s(x_1-x_2)=\rho^{o}_s  e^{i k_s(x_2-x_1)} e^{-\chi_s(x_1-x_2)}.
\label{wick}
\end{equation}
where the phase-phase correlation function is defined as:
\be
\chi_s(x_1-x_2)= \frac{1}{2}\big\langle [\delta\theta_s(x_1,t)- \delta\theta_s(x_2,t)]^2\big\rangle_W.
\label{chidef}
\ee
This can be evaluated using \eq{eqlin} and \eq{w-rho-phi}:
\be 
\chi_s(x)= 
\frac{1}{32\pi^2\,\rho_S^o}\sum_{\nu\mu}\int\!d\omega\,dk\,
\frac{(1-e^{ikx}) \bar \Gamma_{\nu\mu}(k)}{ [\omega-\lambda_\nu^*(k)][\omega-\lambda_\mu(k)]},
\label{chi-Xi-new}
\ee
in terms of the matrix elements
\be
\bar{\Gamma}_{\nu\mu}(k)= P^\dag B_\mu(k)\Gamma(k) B_\nu^\dag(k) P.
\label{defT}
\ee
The vector $P$ is defined as $P=(1,0,0,-1,0,0)^T$ and
the matrices $B_\nu(k)$ are the projectors on the eigenspace
corresponding to the eigenvalue $\lambda_\nu(k)$. 
%In deriving Eq. \eq{defT}, we have used that for any matrix $A$:
%$\langle \mathcal W^\dag(k',\omega') A \mathcal W(k,\omega)\rangle = 
%\text{Tr}[A] \delta(k-k')\delta(\omega-\omega')$.
The frequency integral in \eq{chi-Xi-new} can then be performed closing the integration
contour in the complex plane:
\be
\chi_s(x)= \frac{i}{16 \pi \rho_S^0}\sum_{\nu\mu}
\int dk\,\big(1-e^{ikx} \big)
\frac{\bar{\Gamma}_{\nu\mu}(k)}{\lambda_\nu^*(k)-\lambda_\mu(k)}.
\label{chi-sum}
\ee

\begin{figure}[htbp]
\begin{center}
\includegraphics[width=1\columnwidth,angle=0,clip]{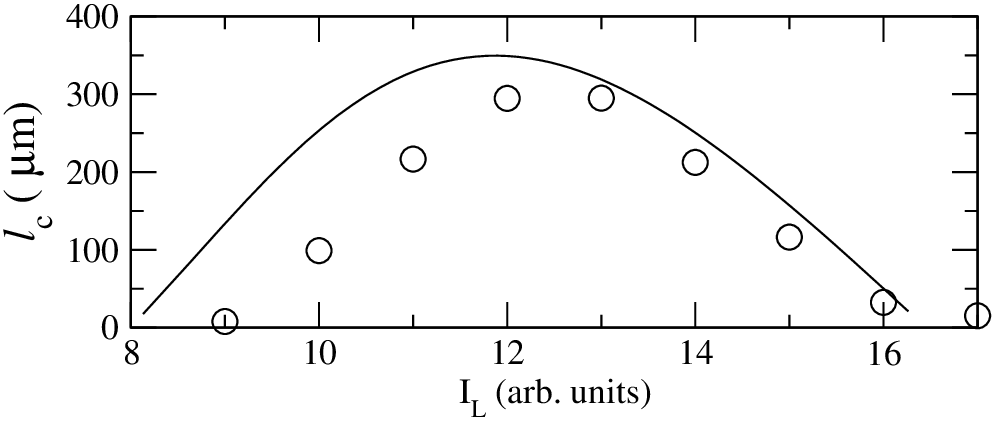}
\caption{
Analytical prediction \eq{lcor} (solid) and QMC data (circles) for the coherence length
 as a function of the pump power.
Same cavity parameters as in Fig.\ref{fig:MF}.
}
\label{fig:coh_length}
\end{center}
\end{figure}

As usual, the large distance properties are determined by the $k\approx 0$ region, 
and the dominant contribution comes from the Goldstone mode, that is for 
$\nu=\mu=G$. 
A simple analytical expression can be obtained by keeping only this
mode, then approximating the smooth function $\bar{\Gamma}_{GG}(k)$ with its
$k=0$ value $\bar{\Gamma}_{GG}^o$ (of the order of the $\gamma$'s), 
and finally extending the integral to infinity.
This results in an asymptotic exponential decrease
\be
g^{(1)}_s(x) \propto e^{-|x|\,/\ell_c},
\ee
with a coherence length $\ell_c$ equal to
\be
\ell_c=\frac{16\,\rho_s^o}{a\,\bar{\Gamma}_{GG}^o}.
\label{lcor}
\ee
In Fig.\ref{fig:coh_length}, $\ell_c$ is plotted
as a function of the pump power. Its behaviour closely follows the one of the
signal density $\rho_s^o$ so that $\ell_c$ goes to $0$ at the edges of the
OPO region, and has its maximum in the middle.
This prediction can be compared to the numerical results of a
full Monte Carlo simulation of the Wigner equation as reported in
Ref.\onlinecite{iac-prb}.
The agreement is good and the small
quantitative discrepancy is mostly due to the three-mode expansion 
performed here, which does not take into account the 
upper polariton.

For a given value of the mean-field interaction energy $g\rho_j$
and all detuning and cavity parameters kept 
fixed, the coherence length has the remarkable scaling $\ell_c \propto 1/g$.
The extremely high value of the excitonic nonlinearity 
implies that the effect of quantum fluctuations is strongly enhanced
as compared to OPOs based on conventional nonlinear optical
media~\cite{gatti,zambrini,threshold,handbook-nonlin-opt}.
In particular, one can observe in Fig.\ref{fig:coh_length} that the predicted coherence length is of
the order of hundred microns for realistic experimental parameters, which means that it should be
 within experimental reach with the current semiconductor technology.

As a final point, it is very instructive to compare the
non-equilibrium formula (\ref{lcor}) with the one for a degenerate 1D
Bose gas of density $n$ at equilibrium. 
The symmetry which is spontaneously broken is in fact the same $U(1)$
in the two cases~\cite{goldstone}.
While at $T=0$ the equilibrium system shows a power law decay of
correlations, an exponential decay is recovered at finite $T$, with a
coherence length given by~\cite{castin}:
\begin{equation}
\ell_{c}^{eq}=\frac{2 n \hbar^2}{k_B T m}
\end{equation}
The dependence on the density is the same in the equilibrium and
non-equilibrium cases, while the role of the temperature $T$ is played
in the non-equilibrium case by the noise associated to damping.
Another, more striking, difference concerns the free boson mass $m$, which is
replaced in the non-equilibrium case by the parameter $a$ defined in
\eq{def_a}, i.e. the inverse of the second derivative of the {\em
  imaginary} part of the Goldstone mode dispersion. 

\section{Conclusions}

In conclusion, we have developed an analytical theory which is able to
describe the long-distance phase coherence of the light emitted by a
one-dimensional optical parametric oscillator above threshold. 
Because of the reduced dimensionality, quantum fluctuations turn out
to be able to destroy the long-range order.
At long distances, the coherence function shows an exponential decay
with a coherence length $\ell_c$ inversely proportional to the
nonlinear coupling constant.   

This result is very general and holds for any kind of one-dimensional
parametrical oscillator, as well as for other examples of
non-equilibrium phase transitions in reduced dimensionality.
We have concentrated our attention on semiconductor photonic wires in
the strong coupling regime simply because such systems appear as
very promising candidates to experimentally investigate this physics.
The extremely large value of the exciton-exciton interactions
corresponds in fact to a value of $\ell_c$ short enough to be within
reach of the current semiconductor technology.  

In addition to their interest for the fundamental physics of pattern
formation and non-equilibrium phase transitions, the conclusions of
the present paper may also have significant consequences for device
applications of OPOs, as quantum fluctuations provide an intrinsic
limitation to the long distance phase coherence that can be
obtained even in the absence of other decoherence and disorder
effects.  

\section{acknowledgements}

Continuous stimulating discussions with C. Ciuti, J. Tignon,
C. Diederichs, and A. Rosso are warmly acknowledged.
This research has been supported financially by the FWO-V projects Nos.  
G.0435.03, G.0115.06, and the Special Research Fund of the University of
Antwerp, BOF NOI UA 2004.
M.W. acknowledges financial support from the FWO-Vlaanderen in the form 
of a ``mandaat  Postdoctoraal Onderzoeker''.


\begin{thebibliography}{99}
\bibitem{cross} M. C. Cross and P. C. Hohenberg, Rev. Mod. Phys. {\bf 65}, 851 (1993).
\bibitem{lugiato-rev}L.A. Lugiato, M. Brambilla and A. Gatti, Adv. At. Mol. Opt. Phys. 40, 229 (1998).
\bibitem{staliunas-book} K. Staliunas, {\em Transverse Patterns in Nonlinear Optical Resonators} (Springer, 2003).
%\bibitem{vaupel} M. Vaupel, A. Ma\^itre, and C. Fabre, Phys. Rev. Lett. {\bf 83}, 5278 (1999).
\bibitem{walls} D. F. Walls and G. J. Milburn, {\em Quantum Optics} (Springer, 1994).
\bibitem{book} L.P. Pitaevskii and S. Stringari, {\sl Bose-Einstein Condensation}, 
Clarendon Press Oxford (2003).

\bibitem{gatti}  A. Gatti, L. Lugiato, Phys. Rev. A {\bf 52}, 1675 (1995). 
\bibitem{iac-prb} I. Carusotto and C. Ciuti, Phys. Rev. B {\bf 72}, 125335 (2005).
\bibitem{noneqstatmech} for an introduction, see 
L. E. Reichel, {\sl A Modern Course in Statistical Physics}, 
University of Texas Press (1980). 
\bibitem{domb}  {\em Phase transitions and critical phenomena}, vol.17 
(Statistical mechanics of driven diffusive systems), eds. C. Domb and J.L. Lebowitz,
Academic, Neq York (1989).
\bibitem{popov} V.N. Popov, {\em Functional
Integrals in Quantum Field Theory and Statistical Physics}, Reidel Dordrecht (1983).
\bibitem{castin} Y. Castin,  J. Phys. IV France, {\bf 116}, 89 (2004) and references therein.


\bibitem{zambrini} R. Zambrini, M. Hoyuelos, A. Gatti,
  P. Colet, L. Lugiato, and M. San Miguel, Phys. Rev. A {\bf 62}, 063801 (2000).
\bibitem{staliunas} K. Staliunas, cond-mat/0001436;
K. Staliunas, Phys. Rev. E {\bf 64}, 066129 (2001).

\bibitem{review1}J. Baumberg and L. Vi\~na (Eds.) Special issue on Microcavities,
[Semicond. Sci. Technol. {\bf 18}, S279 (2003)].
\bibitem{review2} B. Deveaud (Ed.), Special issue on the ``Physics of semiconductors
microcavities'', Phys. Stat. Sol. B {\bf 242}, 2145-2356 (2005) and references therein.

\bibitem{stevenson} R. M. Stevenson,
V. N. Astratov, M. S. Skolnick, D. M. Whittaker, 
M. Emam-Ismail, A. I. Tartakovskii, P. G. Savvidis, J. J. Baumberg, and J. S. Roberts,
Phys. Rev. Lett {\bf 85}, 3680 (2000).
\bibitem{houdre}R. Houdr\'e, C. Weisbuch, R. P. Stanley, U. Oesterle, and M. Ilegems,
Phys. Rev. Lett. {\bf 85}, 2793 (2000).  
\bibitem{dasbach} G. Dasbach, C. Diederichs, J. Tignon, C. Ciuti,
  Ph. Roussignol, C. Delalande, M. Bayer, and A. Forchel, Phys. Rev. B
  {\bf 71}, 161308(R) (2005). 

\bibitem{Panzarini} G. Panzarini and L. C. Andreani, Phys. Rev. B {\bf
  60}, 16799 (1999)

\bibitem{QuantumNoise} C.W. Gardiner and P. Zoller, {\sl Quantum Noise}
  (Springer, 2004).

\bibitem{spinor_inter} T.-L. Ho, Phys. Rev. Lett. {\bf 81}, 742 (1998).

\bibitem{spinor_polar} I. A. Shelykh, Yuri G. Rubo, G. Malpuech,
  D. D. Solnyshkov, and A. Kavokin,  Phys. Rev. Lett. {\bf 97}, 066402
  (2006).

 
\bibitem{iac-superfl} I. Carusotto and C. Ciuti, Phys. Rev. Lett. {\bf
93}, 166401 (2004).
\bibitem{whittaker} D.M. Whittaker, Phys. Rev. B {\bf 71}, 115301
  (2005)
\bibitem{threshold} M. Wouters and I. Carusotto, cond-mat/0607719.
\bibitem{goldstone} M. Wouters and I. Carusotto, cond-mat/0606755.

\bibitem{Kato} T. Kato, {\em Perturbation theory for linear operators} (Springer-Verlag, 1984).

\bibitem{szymanska} While the present paper was under review, similar
  results for a different but related model have appeared in
  M. H. Szymanska, J. Keeling, P. B. Littlewood, Phys. Rev. Lett. {\bf
  96},  230602 (2006) 

\bibitem{baas} A. Baas, J.-Ph. Karr, M. Romanelli, A. Bramati, and
  E. Giacobino, Phys. Rev. Lett. {\bf 96}, 176401 (2006)

\bibitem{handbook-nonlin-opt} R.L. Sutherland, {\em Handbook of nonlinear optics} 
(Marcel Dekker, 2003).


\end{thebibliography}
\end{document}